\documentclass[amsmath,twocolumn,superscriptaddress,prb,aps, longbibliography]{revtex4-1} 
\usepackage{amsthm,amsfonts,graphicx,verbatim, color, physics}
\usepackage{bm}
\usepackage[utf8]{inputenc}
\let\oldmarginpar\marginpar
\renewcommand\marginpar[1]{\-\oldmarginpar[\raggedleft\footnotesize #1]%
{\raggedright\footnotesize #1}}

\newcommand{\be}{\begin{equation}}
\newcommand{\ee}{\end{equation}}
\newcommand{\bea}{\begin{eqnarray}}
\newcommand{\eea}{\end{eqnarray}}

\renewcommand{\epsilon}{\varepsilon}

\DeclareMathOperator{\sgn}{sgn}

\newcommand{\addVR}[1]{{\color{blue}}}

\def\beq{\begin{equation}}
\def\eeq{\end{equation}}
\def\bea{\begin{eqnarray}}
\def\eea{\end{eqnarray}}

\begin{document}

\title{Symmetry breaking and localization in a random Schwinger model with commensuration}
\author{A. A. Akhtar}
\affiliation{Department of Physics, University of California at San Diego,  La Jolla, CA 92093, USA}
\author{Rahul M. Nandkishore}
\affiliation{Department of Physics and Center for Theory of Quantum Matter, University of Colorado at Boulder, Boulder CO 80309, USA}
\author{S. L. Sondhi}
\affiliation{Department of Physics, Princeton University, Princeton, NJ  08544, USA}

\begin{abstract}
We numerically investigate a lattice regularized version of quantum electrodynamics in one spatial dimension (Schwinger model). We work at a density where lattice commensuration effects are important, and preclude analytic solution of the problem by bosonization. We therefore numerically investigate the interplay of confinement, lattice commensuration, and disorder, in the form of a random chemical potential. We begin by pointing out that the ground state at commensurate filling spontaneously breaks the translational symmetry of the lattice. This feature is absent in the conventional lattice regularization, which breaks the relevant symmetry explicitly, but is present in an alternative (symmetric) regularization that we introduce. Remarkably, the spontaneous symmetry breaking survives the addition of a random chemical potential (which explicitly breaks the relevant symmetry) in apparent contradiction of the Imry-Ma theorem, which forbids symmetry breaking in one dimension with this kind of disorder. We identify the long range interaction as the key ingredient enabling the system to evade Imry-Ma constraints. We examine spatially resolved energy level statistics for the disordered system, and demonstrate that the low energy Hilbert space exhibits ergodicity breaking, with level statistics that fail to follow random matrix theory. A careful examination of the structure of low lying excited states reveals that disorder induced localization is responsible for the deviations from random matrix theory, and further reveals that the elementary excitations are charge neutral, and therefore not long range interacting. 
\end{abstract}

\maketitle

\par Quantum electrodynamics in one spatial dimension (also known as the `Schwinger model') has long been an important platform for investigating confinement \cite{Schwinger, Coleman, zittarz, Fishler}. More recently, this model has also proven important in connection with many body localization (MBL), a phenomenon whereby disordered quantum systems can exhibit ergodicity breaking, failing to come to equilibrium even at infinite times \cite{Anderson, Gornyi, BAA,Znidaric, OganesyanHuse,Pal,Imbrie, Nandkishore-2015}. In particular, while MBL traditionally occurs in systems with purely short range interactions, and indeed there are arguments to the effect that long range interactions are incompatible with MBL \cite{Burin, YaoDipoles}, Ref.\onlinecite{NandkishoreSondhi} argued that the interplay of confinement and disorder would cause the disordered Schwinger model to exhibit MBL, despite the long range interaction. Ref.\onlinecite{Mottglass} further pointed out that the ground state of this model realized an elusive and long sought after phase known as the Mott glass \cite{GL, Nattermann}

\par The discussion of the interplay of disorder and confinement in Refs.\onlinecite{NandkishoreSondhi, Mottglass} was formulated in the {\it continuum}. MBL in the continuum is a fairly delicate phenomenon \cite{aleiner2010finite, 2dcontinuum, mullergornyi}, and a lattice formulation of the argument is greatly desired. A lattice formulation is moreover {\it essential} if the scenario of Ref.\onlinecite{NandkishoreSondhi} is to be subjected to numerical tests. At the same time, a reformulation of the problem on a periodic lattice brings in the possibility of {\it commensuration} effects whereby electrons can scatter freely by reciprocal lattice vectors - and these are entirely absent in the continuum analysis. The interplay of confinement, disorder, and {\it commensuration} in lattice regularized versions of the Schwinger model is thus an important open problem. 

\begin{table*}
\begin{tabular}{|c|c|c|}

	\hline 
	\rule[1ex]{0pt}{3ex}  & Open Boundary Conditions (OBC) & Periodic Boundary Conditions (PBC) \\ 
	
	\hline
	\rule[1ex]{0pt}{3ex} $H_{hop}$ & $\sum_{n=0}^{N-2} \left( \sigma_{n}^{+}\sigma_{n+1}^{-} + \sigma_{n}^{-}\sigma_{n+1}^{+} \right)$ & $ \sum_{n=0}^{N-1} \left( \sigma_{n}^{+}\sigma_{n+1}^{-} + \sigma_{n}^{-}\sigma_{n+1}^{+} \right) $ \\
	
	\hline 
	\rule[1ex]{0pt}{3ex} $H_{int}$ & $\sum_{n=0}^{N-2}(E_{n})^{2}$ & $ -\sum_{j<i}\rho(i)\rho(j)\sin \left( \frac{\pi}{N} (i-j) \right) $ \\
	
	\hline
	\rule[1ex]{0pt}{3ex} Conventional & \shortstack{$E_{n}=\frac{1}{4}\sum_{k=0}^{N-1} \sgn (n+\frac{1}{2}-k) \sigma_{k}^{z} +$ \\ $\frac{1}{4}((-1)^{n}+1)$} & $\rho(n) = \frac{1}{2} (\sigma_{n}^{z}+(-1)^{n}) $\\ 
	
	\hline 
	\rule[1ex]{0pt}{3ex} Symmetric & $E_{n}=\frac{1}{4}\sum_{k=0}^{N-1} \sgn (n+\frac{1}{2}-k) \sigma_{k}^{z}$ & $\rho(n) = \frac{1}{2}\sigma_{n}^{z} $ \\
	
	\hline 
\end{tabular}
\caption{We detail the Hamiltonians for open and periodic boundary conditions respectively. The Hamiltonians are obtained by starting from a lattice regularized Schwinger model, performing a Jordan-Wigner transformation, and applying the Gauss's law constraint. The row marked `conventional' indicates the conventional Kogut-Susskind staggered potential regularization with unit cell of length two. This regularization is based on a picture of the Schwinger model as a two component plasma of electrons and positrons. The row labeled `symmetric' indicates results for the alternative `symmetric' regularization that we introduced, with unit cell of length one. This regularization is based on a picture of a one component plasma of electrons moving on a neutralizing `jellium' background, which is more natural from a condensed matter perspective. Following standard notation, $\sigma^{z}_{n}$ is the $z$-Pauli matrix for the Hilbert space on site $n$, and $\sigma^{\pm}_{n} \equiv \frac{1}{2}(\sigma^{x}_{n} \pm i\sigma^{y}_{n})$ are the raising (lowering) operators on site $n$. \label{tab: Hamiltonian}}
\end{table*}

\par In this work, we investigate the interplay of confinement, disorder, and lattice commensuration effects in a disordered, lattice regularized Schwinger model, using numerical exact diagonalization (ED) and density matrix renormalization group (DMRG) techniques. The presence of strong lattice commensuration precludes analytic solution of the problem by bosonization, as in Ref. \onlinecite{NandkishoreSondhi}, and necessitates a numerical approach. We discover that the ground state of the lattice Schwinger model at commensurate filling spontaneously breaks translation symmetry. This feature is absent in the conventional Kogut-Susskind \cite{Kogut-Susskind} lattice regularization of the Schwinger model, which breaks translation symmetry explicitly. It is however crucial in an alternative (fully symmetric) regularization that we introduce. We demonstrate that the broken symmetry phase survives the addition of a random chemical potential, which itself explicitly breaks lattice translation symmetry (`random field disorder' in the statistical mechanics jargon). This result flies in the face of the Imry-Ma theorem \cite{ImryMa}, which states that symmetry breaking in one spatial dimension is unstable to random field disorder. We identify the long range interaction as the key ingredient enabling the system to evade Imry-Ma constraints. We then examine the energy resolved level statistics and hence show that the low energy Hilbert space exhibits ergodicity breaking. While ergodicity breaking could come from either integrability or localization, a careful DMRG based examination of the structure of low lying excited states indeed reveals disorder induced localization, as was predicted (in the absence of commensuration) by Ref. \onlinecite{NandkishoreSondhi}. The DMRG analysis also reveals that the elementary excitations are charge neutral on average, and therefore lack long range interactions. 

\par This paper is structured as follows. In Sec.\ref{sec:regular} we discuss the regularization of the Schwinger model on a lattice. In Sec.\ref{sec: sym} we discuss spontaneous symmetry breaking first in the clean and then the disordered Schwinger model. In Sec.\ref{sec: erg} we discuss level statistics and ergodicity breaking in the low energy Hilbert space. Finally, in Sec.\ref{sec: localization} we discuss localization in the low lying excited states. We conclude with a discussion of our results. 

\section{Lattice regularization of the Schwinger model}
\label{sec:regular}

\par The Schwinger model is a 1+1 dimensional model for quantum electrodynamics, describing the interaction of a Dirac field $\Psi$ (charge $e$) with an electric field $E$ that is conjugate to a vector potential $A^{\mu}$. The continuum action (including a random chemical potential) takes the form

\begin{equation}
	\mathcal{L} = \bar{\Psi}(i\gamma^{\mu}\partial_{\mu} - e\gamma^{\mu}A_{\mu} + \mu(x))\Psi - \frac{1}{4}F_{\mu \nu}F^{\mu \nu},
\end{equation} 

where the chemical potential depends only on spatial position. In the temporal gauge ($A_{0} = 0$) the continuum Hamiltonian takes the form:

\begin{equation}
	H = \int dx \left(\bar{\Psi}(-i\gamma^{1}\partial_{1} + e\gamma^{1}A_{1} + \mu(x))\Psi + \frac{1}{2}E^{2}\right),
\end{equation}

where $E = - \partial_0 A_1$ is the electric field conjugate to the vector potential. The electric field is fixed by Gauss's law $\partial_1 E = e \bar{\Psi} \gamma^0 \Psi$ up to a constant of integration, corresponding to a uniform background field. We set the background field to zero in all that follows.

\par We wish to regularize this model on a lattice. We wish to work at half filling (one electron for every two lattice sites), so as to maximize lattice commensuration effects. The Kogut-Susskind staggered potential formulation \cite{Kogut-Susskind} is the conventional way to do this. A brief summary of this approach can be found in Ref.\onlinecite{Banuls}. In essence one introduces fermion fields $\phi$ which sit on lattice sites, and gauge fields that sit on links. The conjugate variable to the gauge field is a coordinate $L_{n},$ which is related to the electric field by $eL_{n} = E_{n}$, where $E_{n}$ is the electric field between sites $n$ and $n+1$. If the gauge field is taken to be compact, then Gauss's law takes the form

\begin{equation}
	L_{n} - L_{n-1} = \phi^{\dagger}_{n}\phi_{n} - \frac{1}{2} [ 1 - (-1)^{n} ]. \label{eq: conventionalGauss}
\end{equation}   

\par In this conventional Kogut-Susskind regularization, the ground-state expectation value $<\phi_{n}^{\dagger}\phi_{n}> = 0$ (1) for $n$ even (odd). The regularization thus explicitly breaks the translation symmetry of the lattice, giving rise to a staggered potential with a two site unit cell. In essence, the regularization places electrons on odd sites, and positrons on even sites, leading to charge neutrality on average. Underlying this regularization is a picture of a two component plasma, where electrons and positrons are simultaneously present with equal density. This model may then be recast as a spin Hamiltonian by performing a Jordan-Wigner transformation and applying the Gauss's law constraint. In this manner one obtains a Hamiltonian of the form 

\begin{equation}
	H = x H_{hop} + \lambda H_{int} + \frac{1}{2}\sum_{n=0}^{N-1}V(n)\sigma_{n}^{z}  \label{eq: basic Hamiltonian}
\end{equation}

where $V(n)$ comes from the random chemical potential and is chosen from the uniform distribution $( - \theta, \theta)$, with $n \in 0...N-1$, where $N$ is the system size. Meanwhile the specific forms of $H_{hop}$ and $H_{int}$ depend on the choice of boundary conditions and are tabulated in Table \ref{tab: Hamiltonian}. The parameters $x, \lambda, \theta$ characterize the hopping, interaction, and disorder strengths of Hamiltonian, respectively. 

\par It is illuminating instead to consider an alternative `symmetric' regularization with a one site unit cell, motivated by an alternative picture of the Schwinger model as a one component plasma, where electrons move on a neutralizing `jellium' background. This symmetric model is obtained by removing the staggered potential and modifying the Gauss's law to
 
\begin{equation}
	L_{n} - L_{n-1} =  \phi^{\dagger}_{n}\phi_{n} \label{eq: symmetricGauss}
\end{equation}

\par With this regularization $\langle \phi_{n}^{\dagger}\phi_{n} \rangle = 0 \, \forall n$, and the unit cell consists of a single lattice site. Moving to a spin model as before, one obtains again a Hamiltonian of the form (\ref{eq: basic Hamiltonian}), where however the `interaction' Hamiltonian takes the distinct form tabulated in Table.\ref{tab: Hamiltonian}. The two distinct choices of Gauss's law (\ref{eq: conventionalGauss}, \ref{eq: symmetricGauss}) correspond to two distinct lattice regularizations of the same continuum model - however, they have (apparently) very different properties. In particular, the conventional regularization has a unit cell of period two, while the `symmetric' regularization has a unit cell of period one. Nevertheless, as we shall see, they have the same long distance physics, since the model with `symmetric' regularization spontaneously breaks symmetry so that the states have a unit cell of period two. 

\par The Hamiltonian given by Eq.\ref{eq: basic Hamiltonian} together with Table \ref{tab: Hamiltonian} constitutes the basic model we study in this work. To simplify our analysis, we work throughout in the total spin-zero sector i.e. $(\sum_{n} \sigma^{z}_{n}) \ket{j} = 0$ for the basis states $\{ \ket{j} \}$. This is equivalent to working in a system with zero net charge.

\section{Symmetry breaking in disordered Schwinger model}
\label{sec: sym}
\par The Schwinger model may be regularized on a lattice in one of two distinct ways, as we have discussed. The conventional regularization is based on a picture of a two component plasma of electrons and positrons and has a two site unit cell. An alternative regularization may also be used however, based on a picture of electrons moving in a uniform positively charged background, and this regularization leads to a {\it Hamiltonian} with only a one site unit cell. Nevertheless, as we will demonstrate, both regularizations encode the same physics.

\par To develop some intuition for why the low energy physics may be insensitive to the choice of regularization, recall that the classical one component plasma develops long range crystalline order at any temperature \cite{Baxter, Baus}, while the two component plasma does not \cite{Choquard}. One may thus conjecture that the {\it quantum} regularization based on a model of a one component plasma may {\it spontaneously} break symmetry, exhibiting crystalline order with a two site unit cell. This would then lead to results analogous to those obtained with the conventional regularization, which employs a {\it Hamiltonian} with a two site unit cell. 

\par To test this scenario, we first employ exact diagonalization on systems with $N=12$ spins. In the interests of pedagogy we provide below some details of the numerical procedure.  A natural basis for computation is the basis defined by the local $\sigma^{z}$ operators. In this basis, the action of $\sigma_{n}^{z}$, $\sigma_{n}^{\pm}$ is easily programmed. To minimize memory-use, we represent each vector of the fixed total-spin subspace as a bit-vector. It's clear the value of $\frac{1}{2}\sum_{n}\sigma_{n}^{z}$ fixes the number of zeros and ones, and so we can write a simple recursive method to generate the basis states without repetition. We order the basis states in alphanumeric order, thus $0$ (1) is mapped to $\ket{1}$ ($\ket{2}$) eigenvector of $\sigma^{z}$. We can now construct the Hamiltonian directly in this basis as an $n \times n$, sparse matrix, where $n = \binom{N}{N/2}$ in the total-spin zero case. We construct the Hamiltonian in $O(N n\lg n)$ time, making use of the fact that any bitvector $\ket{j}$, there are at most $O(N)$ $\ket{i}$ such that $\bra{i}H\ket{j}$ is non-zero\footnote{If one uses a dictionary and defines a straightforward hash function, one could reduce the time complexity to $O(Nn)$, but construction only happens once per realization and is not a bottleneck of the algorithm. }. For each bitvector $\ket{j}$, we compute the possible $\ket{i}$ such that $\bra{i}H\ket{j}$ is non-zero, and to determine what row in the matrix each $\ket{i}$ corresponds to takes $\log n$ time using a simple binary search in the basis. Since we store the Hamiltonian as a sparse matrix, this procedure only uses $O(Nn)$ memory, which is far superior to actually constructing the full $n \times n$ matrix. Furthermore, this is much more efficient than constructing the full $2^{N} \times 2^{N}$ Hamiltonian first, then projecting into the total-spin zero subspace.

\par There are a number of additional tricks that we employ to speed up the implementation. For the largest systems we consider, we only compute a small subset of the spectrum, using Lanczos. When we consider multiple disorder realizations, we speed up re-calculating the Hamiltonian for each sample given a set of parameters by subtracting off the old disorder and adding the new disorder in. This hastens the implementation because we aren't constructing a new Hamiltonian for each sample, just modifying the existing one in memory. Finally, the calculation is extensively parallelized.  
 
\par
\begin{figure*}
	\includegraphics[width = 6.5cm]{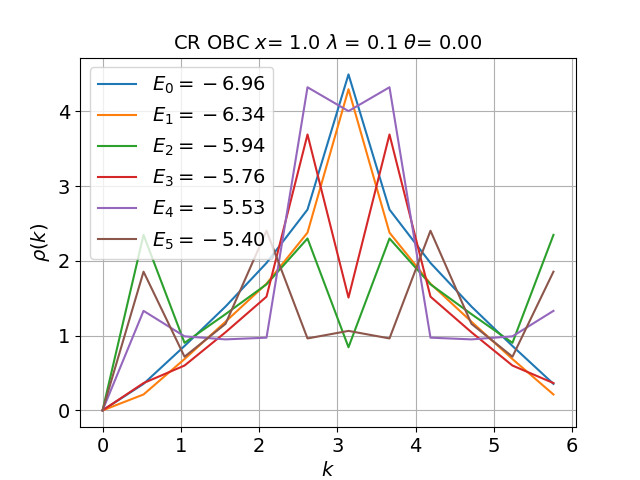}
	\includegraphics[width = 6.5cm]{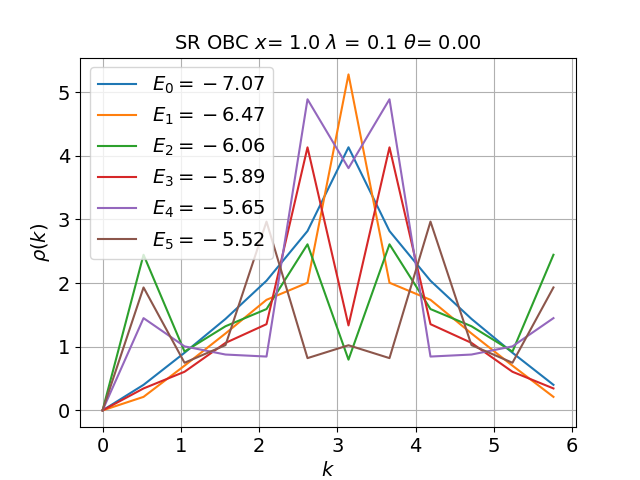}
	\includegraphics[width = 6.5cm]{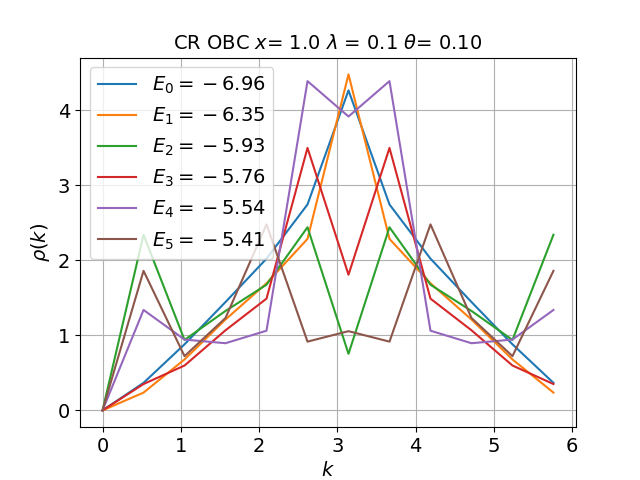}
	\includegraphics[width = 6.5cm]{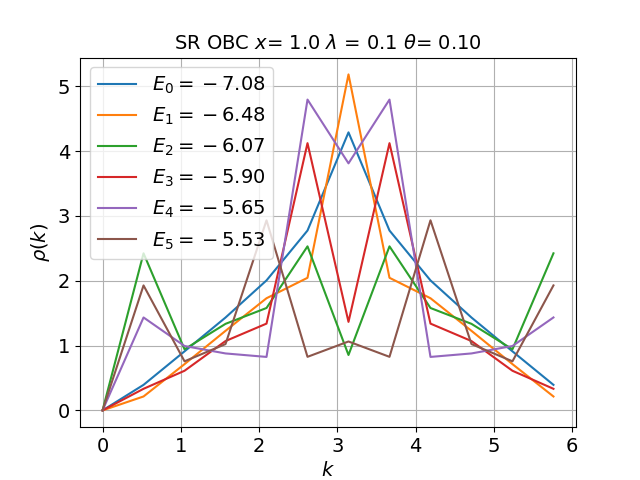}
	\includegraphics[width = 6.5cm]{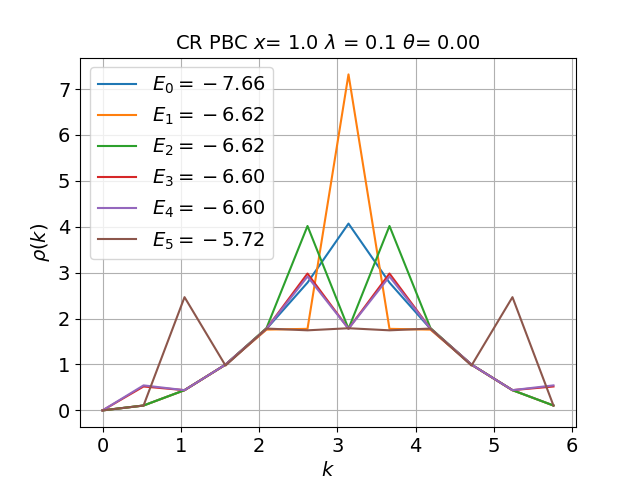}
	\includegraphics[width = 6.5cm]{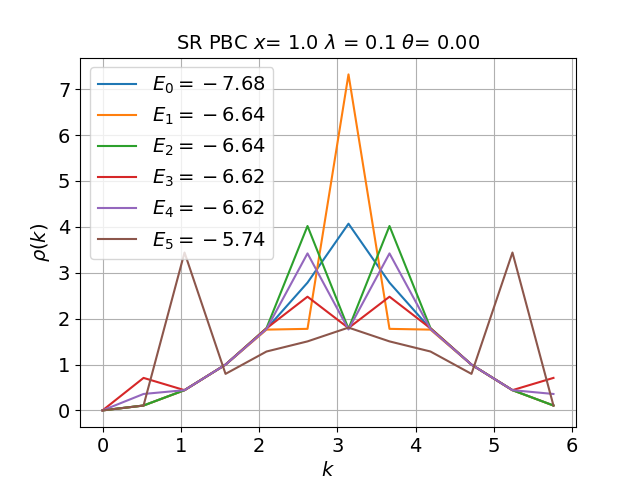}
	\includegraphics[width = 6.5cm]{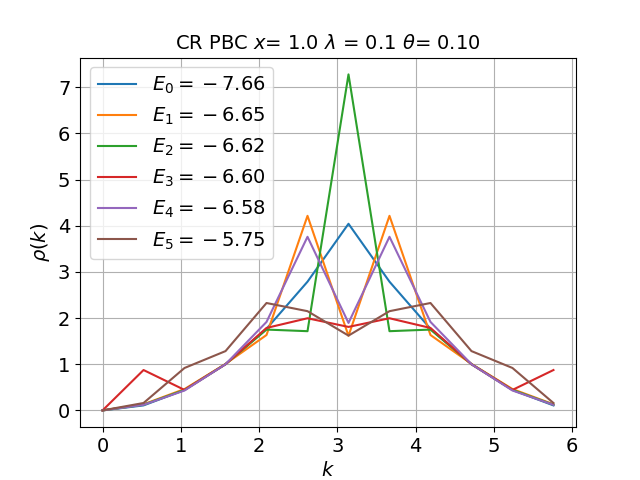}
	\includegraphics[width = 6.5cm]{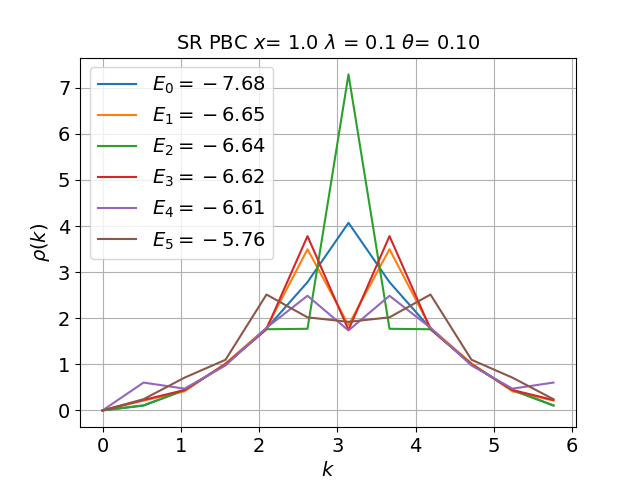}
	\caption{Fourier transform of the density density correlation function $\langle \sigma^z_x \sigma^z_{x+r} \rangle$ in the ground state $E_0$ and first five excited states, as obtained using exact diagonalization on a system of twelve spins. The plots in the left hand column are with the conventional regularization (CR), based on a picture of a two component plasma, and with a two site unit cell. The plots in the right hand column are for the alternative `symmetric' regularization (SR), based on a picture of a one component plasma, and with a one site unit cell. The first two rows show data obtained with open boundary conditions (OBC), and the last two rows show periodic boundary conditions (PBC). The labels $x, \lambda, \theta$ denote the strength of the kinetic, interaction and disorder terms, respectively. Remarkably, the choice of lattice regularization (and the presence or absence of disorder) makes little difference to the density correlation function, which has structure around $k = \pi$ indicating that the {\it ground} state has a two site unit cell. \label{fig: fig1} }
\end{figure*}

 Having diagonalized the Hamiltonian in this way, we look at the Fourier transform of the density density correlation functions $\langle \sigma^z_x \sigma^z_{x+r}\rangle$ in the ground and first five excited states. The results are shown in Fig.\ref{fig: fig1}. Note that while the results depend on choice of boundary conditions, there is little dependence on the choice of regularization. In particular, regardless of regularization there is a peak at momentum $k \approx \pi$ indicating crystalline order with a two site unit cell. This supports the basic scenario of spontaneous symmetry breaking in the `symmetric' regularization that we have introduced. The (considerable) additional structure, particularly in the Fourier transform of the excited states, is we believe a finite size effect. 

\begin{figure*}
	\centering
	\includegraphics[width = 6.5cm]{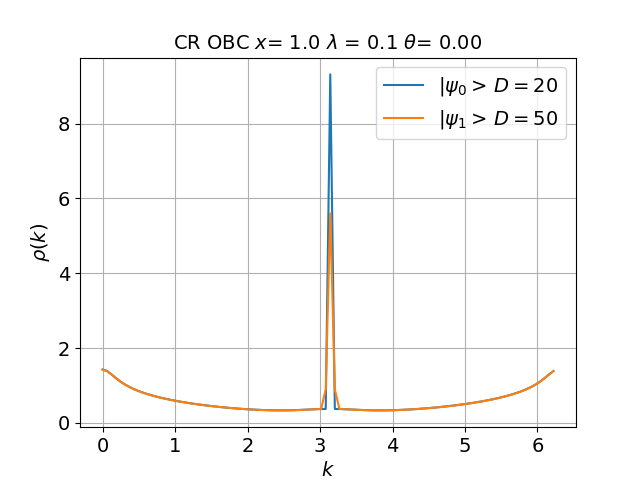}
	\includegraphics[width = 6.5cm]{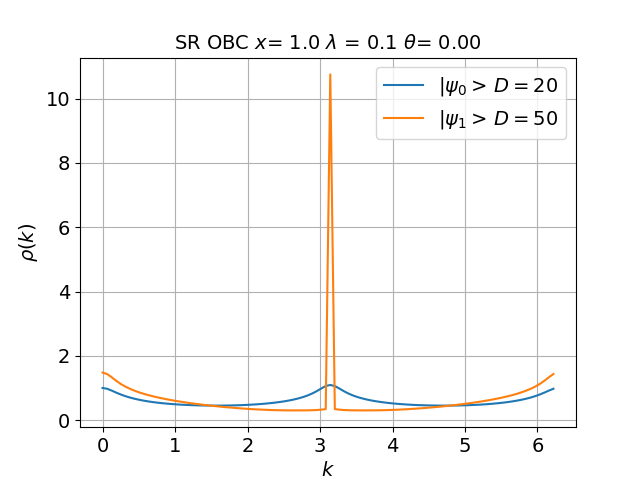}
	\includegraphics[width = 6.5cm]{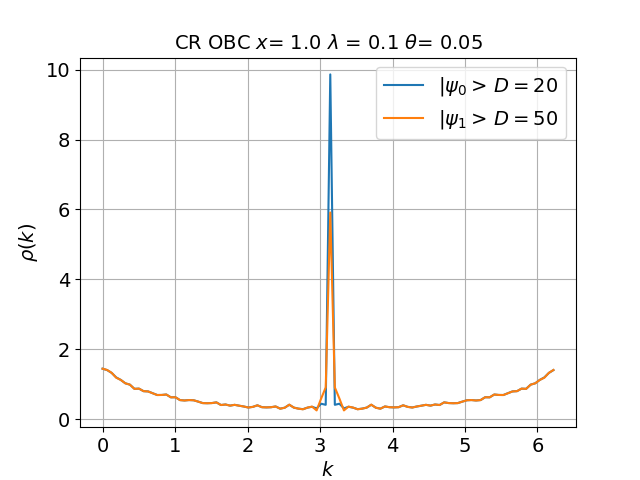}
	\includegraphics[width = 6.5cm]{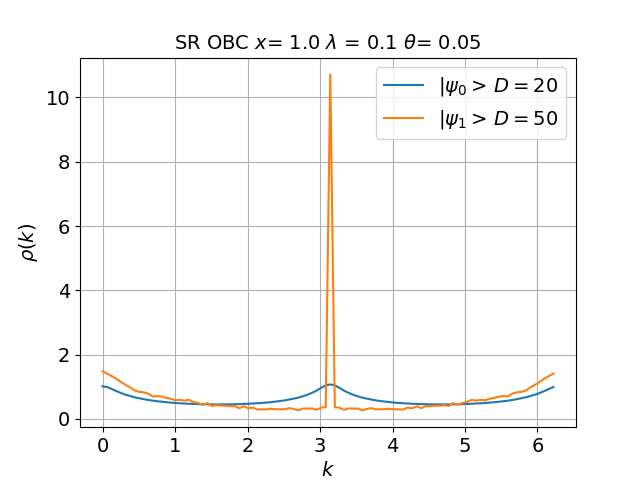}
	\includegraphics[width = 6.5cm]{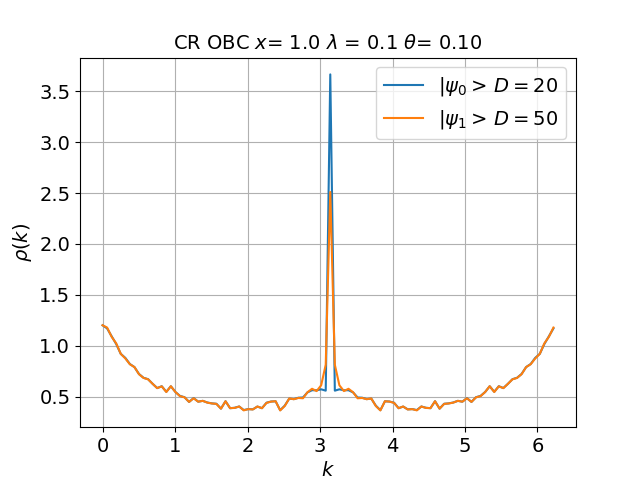}
	\includegraphics[width = 6.5cm]{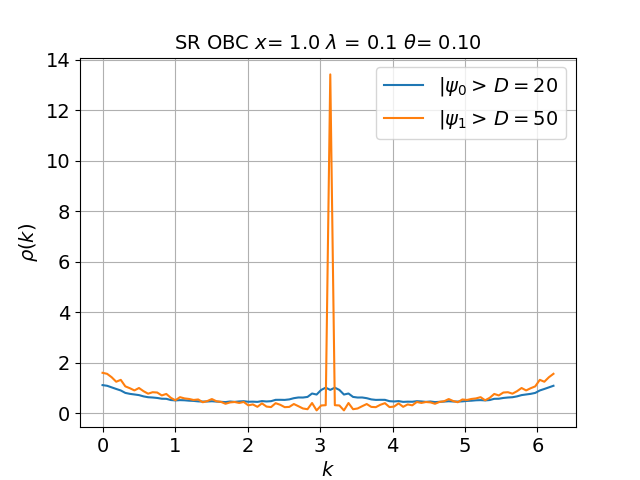}
	\caption{Fourier-transformed density-density correlations in the ground and first excited state, obtained using DMRG for a chain with $N=100$ spins and open boundary conditions. The first column corresponds to the conventional regularization, the second column to symmetric regularization. Bond dimension $D$ is indicated on the plots, as is interaction strength $\lambda$ and disorder strength $\theta$. The disorder strength $\theta = 0.0, 0.5, 0.1$ going top down. The peak at $\pi$ indicates density structure with a two site unit cell, regardless of choice of regularization or addition of disorder. 
	\label{fig: fig2}}
\end{figure*}

\par Cleaner results may be obtained by enlarging the system size. To do this we employ DMRG techniques (following Ref. \onlinecite{Banuls}). This allows us to treat chains of up to $N=100$ spins, although we are limited to systems with open boundary conditions, and are not able to easily access excited states. The results obtained in this manner are plotted in Fig. \ref{fig: fig2}. Note that expanding system size leads to significantly cleaner results. Note also the very clear peak at momentum $k=\pi$, indicating density structure with a two site unit cell, in both the conventional and symmetric lattice regularizations. This further supports our basic thesis that both the conventional and symmetric lattice regularizations encode the same physics, since the symmetric regularization spontaneously breaks symmetry until the system is no more symmetric than with conventional regularization. Why in symmetric regularization the peak is so much sharper in the first excited state than in the ground state is however a mystery. 

\par We now comment on a remarkable feature hidden in Fig.\ref{fig: fig1}, \ref{fig: fig2}. Namely, the random scalar potential ($V(n)$ in the lattice regularization) {\it explicitly} breaks translation symmetry. That is, it constitutes `random field' disorder, in the language of the Imry-Ma theorem \cite{ImryMa}. However, spontaneous symmetry breaking is supposed to be {\it unstable} to random field disorder in one spatial dimension. How then does the system with the `symmetric' regularization manage to develop a two site unit cell, even in the presence of disorder? 

\par The resolution to this apparent puzzle relies sensitively on the existence of a long range interaction in the Schwinger model. In the interests of pedagogy, however, we first review how the basics of the Imry-Ma argument. The Imry-Ma argument proceeds by noting that forming domains of size $L$ that locally align with `disorder' should gain an energy $\sim \theta L^{d/2}$ based on central limiting arguments, where $\theta$ is the typical disorder strength and $d$ is the spatial dimension. Meanwhile, the energy cost of domain walls (in short range interacting systems) is of order $L^{d-1}$. For $d < 2$ and sufficiently large $L$, the system can then always gain energy by creating domains that locally align with the disorder - and this destroys long range order. 

\par To understand how the disordered Schwinger model evades this argument, we must realize that the `spontaneous symmetry breaking' of the Schwinger model (in its symmetric regularization) corresponds to {\it charge density wave} formation with period two (in effect, Wigner crystallization). A domain wall in this charge density wave is a soliton where the system switches from having charge preferentially on odd numbered sites to having charge preferentially on even numbered sites, or vice versa. However, this kind of `phase slip' necessarily binds a charge. If the phase slip is accomplished by leaving two successive sites unoccupied, then it binds a positive charge; if it is accomplished by placing electrons on two successive sites, it binds a negative charge. However, charges in the Schwinger model experience a {\it long range} interactions. Thus, creating a domain of size $L$ involves creating two domain walls at separation $L$, and this carries an energy cost $\sim L$, which for large $L$ overwhelms the $\sim L^{1/2}$ energy gain from locally aligning with disorder. The Imry-Ma argument is thus evaded. The system can {\it locally} deform, making {\it small} domains that align with disorder in regions where the disorder is small, but the domain walls are always bound together by the long range (confining) interaction, and thus long range order survives. The robustness of long range order to disorder is quite apparent in our numerical results. 

\par As a parenthetical remark, we note that in the symmetric regularization, sites either carry charge $-e/2$ if an electron is present or $+e/2$ if an electron is absent. A domain wall consisting of two consecutive occupied/unoccupied sites thus carries a net {\it half integer} charge. This half integer charge is similar in origin to the celebrated Su-Schrieffer-Heeger model (Ref. \onlinecite{SSH}), and may be understood as the polarization charge corresponding to the shift of an electron by half a unit cell. It implies that the domain wall solitons in the symmetric regularization may be viewed as {\it fractionally charged quarks}. Of course, these `quarks' are confined into `mesons' (dipoles) by the long range interaction.

\section{Ergodicity breaking in low energy Hilbert space}
\label{sec: erg}

\par We now search for ergodicity breaking in the low energy Hilbert space of the disordered Schwinger model, using as our principal diagnostic the level statistics ratio \cite{Huse-Pal}. The level statistics ratio $\langle\tilde{r}\rangle$ is extracted from the spectrum of eigenvalues as follows:

\begin{enumerate}
	\item Compute the sorted spectrum of eigenenergies $\{ E_{i} \}$, with $E_{0} \leq E_{1} ... \leq E_{N-1}$. 
	\item Define the gap vector $S$ by $S_{i} = E_{i + 1} - E_{i}$. 
	\item Define the vector $\tilde{r}$ by $\tilde{r}_{i} = \min (S_{i + 1}, S_{i}) / \max (S_{i + 1}, S_{i})$. 
	\item The level statistics ratio is the mean of $\tilde{r}$.
\end{enumerate} 

\par
\begin{figure}
	\centering
	\includegraphics[width= 8cm]{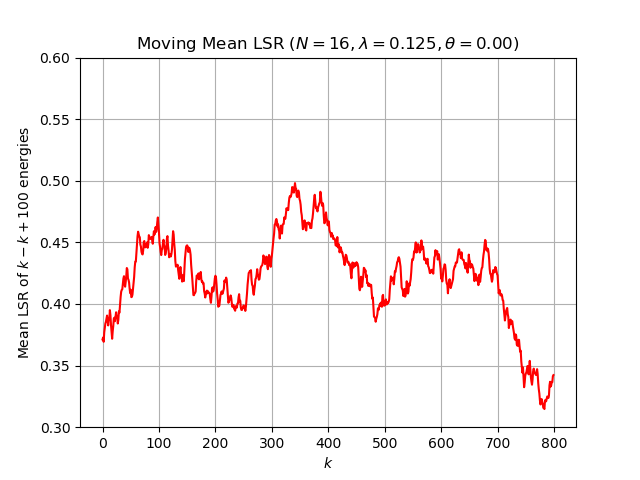}
	\includegraphics[width= 8cm]{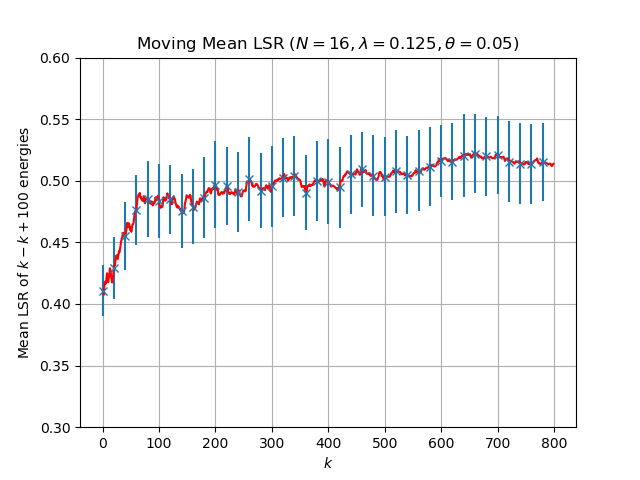} 
	\caption{Energy resolved level statistics ratios for the disordered Schwinger model, obtained using numerical exact diagonalization on $N=16$ spins. The conventional regularization is employed, and $x=1.0, \lambda = 0.125,\theta$, are the energy scales for the kinetic, interaction and disorder pieces of the Hamiltonian respectively. Disordered data has been averaged over one hundred disorder realizations (from which we estimate the error bars shown in blue). Note the absence of level repulsion at low energies. \label{fig: fig3}}
\end{figure}

 The level statistics ratio has a number of salient properties. Firstly, the level statistics ratio is independent of the basis, since it is only a function of the eigenvalues. The level statistics ratio is also invariant under scaling and shifting the spectrum. Finally, the level statistics ratio is governed by random matrix theory in ergodic phases, and by the Poisson distribution in non-ergodic (localized or integrable) phases. For a system in the orthogonal class (as here) the characteristic values are $\langle \tilde{r} \rangle_{GOE} = 0.5314$ in the ergodic phase and $\langle \tilde{r}\rangle_{Poisson} = 0.3836$ in a localized or integrable phase. We explore the level statistics ratio of the Schwinger model using numerical exact diagonalization on a chain with $N=16$ spins. Since we are particularly interested in the low energy Hilbert space, we examine the energy resolved level statistics ratio $\langle r \rangle(E)$ obtained by averaging over energy windows containing one hundred eigenvalues at a time. The results obtained in this manner are shown in Fig.\ref{fig: fig3}. 
 
\par We note first that in the absence of disorder $\theta = 0$, the level statistics are quite far from ergodic. We speculate this may be because of proximity to an integrable point (the clean continuum Schwinger model). Upon turning on disorder, the high energy sector of the Hilbert space starts to show ergodic level statistics. However, the level statistics ratio in the low lying Hilbert space exhibits clear signs of ergodicity breaking, with a level statistics ratio close to the Poisson value. 

\par
\begin{figure}
	\includegraphics[width= 6cm]{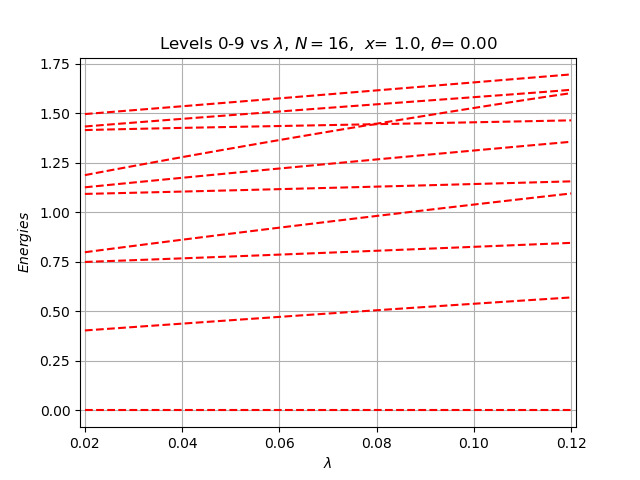}
	\includegraphics[width= 6cm]{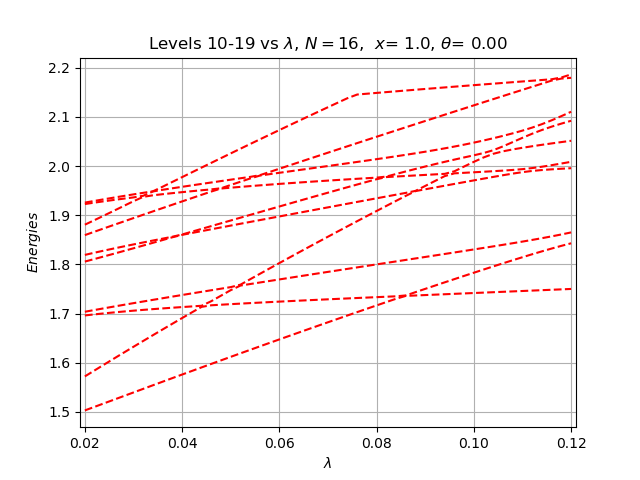}
	\caption{Eigenenergies of first twenty states as obtained by exact diagonalization on a system of sixteen spins. Note the clear level crossings in the clean system, indicative of residual integrability. \label{fig: fig4}}
\end{figure}

 How should the ergodicity breaking in the low energy Hilbert space be understood? It is tempting to attribute this to localization, but the calculations of Ref.\onlinecite{NandkishoreSondhi}, with our parameter values, suggest a localization length of order twenty sites, which is not expected to be accessible in a sixteen site simulation. Furthermore, there is a more prosaic explanation - namely a residual {\it integrability} of the model. (The lattice Schwinger model is not integrable as far as we are aware, but the clean continuum Schwinger model is integrable, and `proximity' to this integrable point may contaminate our results given the limitations of finite size). To test for this possibility, we have tracked the energies of low lying excited states as a function of interaction strength in Fig. \ref{fig: fig4}. We find clear signatures of level crossings in the low lying energy spectrum of the clean system, indicating that the low energy part of Hilbert space is behaving in an integrable manner at least in the finite sizes that we can access. Level statistics alone therefore cannot discriminate between localization and integrability as potential explanations of the non-ergodic level statistics in the low energy Hilbert space. 

\section{Localization in the low energy Hilbert space}
\label{sec: localization}
\par We have observed that level statistics are indicative of ergodicity breaking in the low energy Hilbert space of the disordered lattice Schwinger model, but cannot discriminate between localization and a residual integrability in this finite size system. We therefore directly examine the {\it eigenstates}. Since the localization lengths are estimated to be quite long \cite{NandkishoreSondhi} ($\xi \approx 20$ for the parameter values under consideration \footnote{The parameter values, meanwhile, are chosen to ensure that we are in a regime where bosonization is well controlled}). we employ DMRG, which allows us to treat systems with $N=100$ spins. We do this by first constructing a matrix product operator (MPO) representation of the Schwinger Hamiltonian. Then, for a given bond dimension, we construct a random matrix product state (MPS), and then use the iterative ground state search algorithm that works by updating the tensors on each site of the MPS. We sweep through the tensors site by site, until the algorithm converges on some value for the energy. A detailed explanation of the algorithm is found in Ref. \onlinecite{MPS-DMRG}. Diagonalizing the network at each iterative step is done using Lanczos' algorithm. The bond dimension of the Schwinger Hamiltonian MPO is $D = 5$. The ground state is represented by an MPS with $D=20$. Excited states are calculated as in Ref.\onlinecite{Banuls}, by first finding the ground state and then adding a penalty term to the original Hamiltonian, i.e.
 
\begin{equation}
	H \rightarrow H + w\ket{\psi_{0}}\bra{\psi_{0}}
\end{equation}

where $w$ is a large positive constant ($w >> E_{1} - E_{0}$). Minimizing this Hamiltonian using the same iterative ground search algorithm gives the next excited state. For the first excited state, we use a larger bond dimension $D=50$. Unfortunately, iterating this process to find higher energy excited states is numerically very costly, so we are limited to the ground state and first excited state. We are also limited to open boundary conditions. 

\begin{figure*}
	\centering
	\includegraphics[width=7cm]{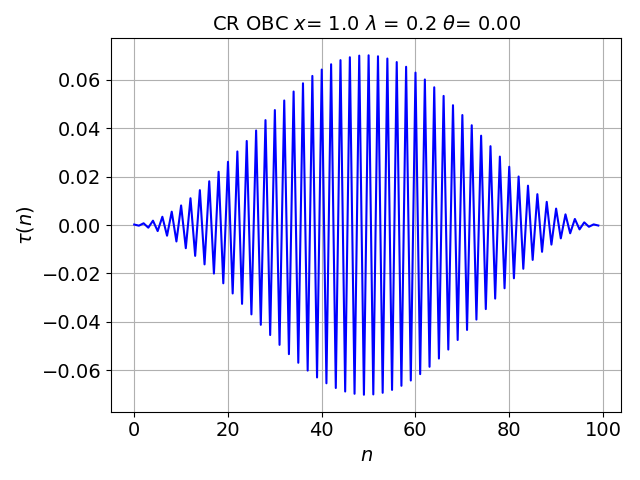}
	\includegraphics[width=7cm]{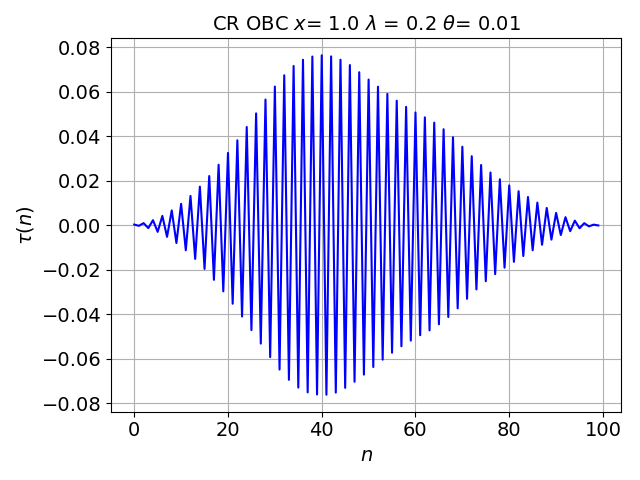}
	\includegraphics[width=7cm]{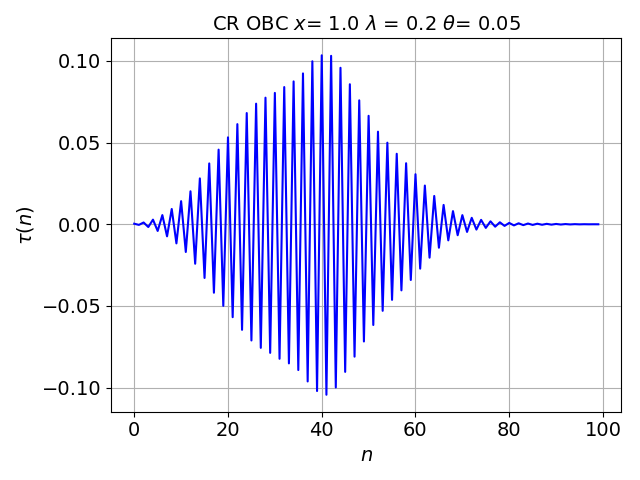}
	\includegraphics[width=7cm]{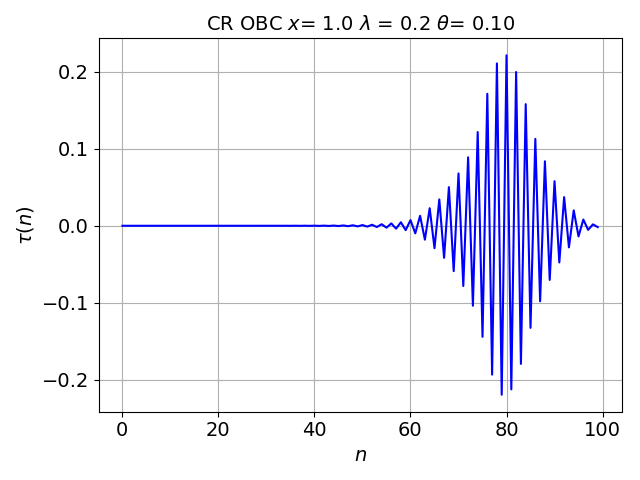}
	\caption{Spatial profile of density imbalance (Eq:\ref{imbalance}) between ground state and first excited state for a system of $N=100$ spins, conventional regularization, and open boundary conditions treated using DMRG. Here $ x = 1.0, \lambda = 0.2$ and $\theta=0,0,01,0.05, 0.1$ indicate the strength of the kinetic, interaction, and disorder terms respectively. Note that the results are clearly indicative of the first excited state (of a disordered system) containing a localized excitation, the localization length of which decreases as disorder strength is increased. Note also that the localization length is too large to be accessible with exact diagonalization (consistent with analytic predictions \cite{NandkishoreSondhi}) thus necessitating DMRG. Note also that the elementary excitations shown here are charge neutral (the density difference is equally likely to be positive or negative), consistent with general expectations.  \label{fig: fig5} }
\end{figure*}

\par To probe localization, we examine the density imbalance between the ground $\ket{\psi_{0}}$ and first excited $\ket{\psi_{1}}$ states. Specifically, we extract 

\begin{equation}
	\tau(n) = \bra{\psi_{1}} \sigma_{n}^{z} \ket{\psi_{1}} - \bra{\psi_{0}} \sigma_{n}^{z} \ket{\psi_{0}} \quad n \in \{ 0...N-1 \} \label{imbalance}
\end{equation}

\par
\begin{figure*}
	\centering
	\includegraphics[width=6cm]{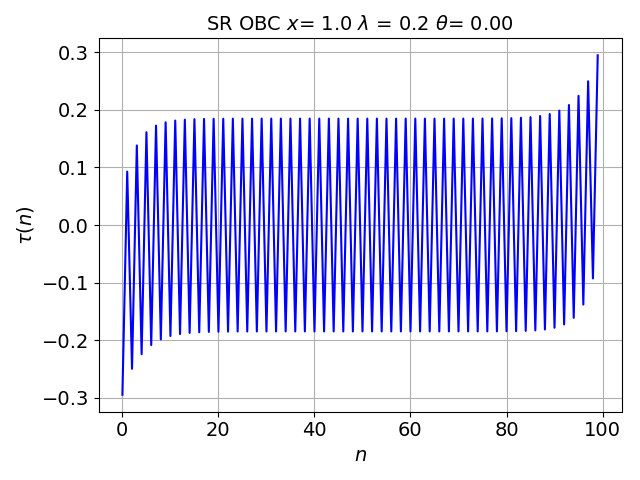}
	\includegraphics[width=6cm]{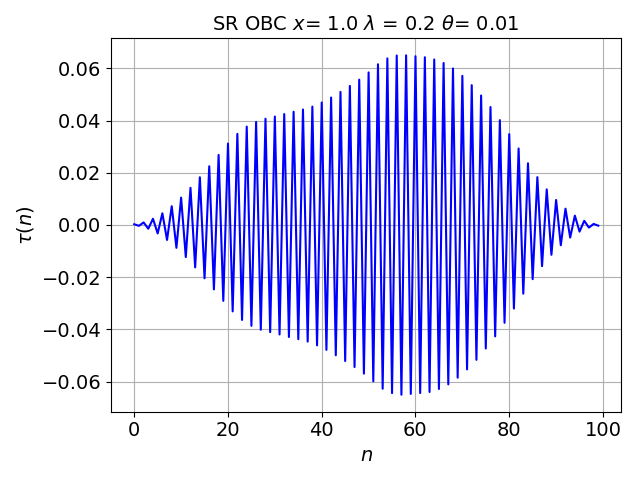}
	\includegraphics[width=6cm]{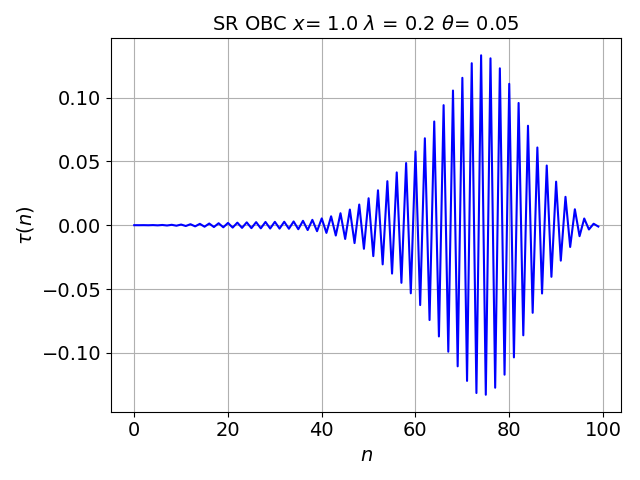}
	\includegraphics[width=6cm]{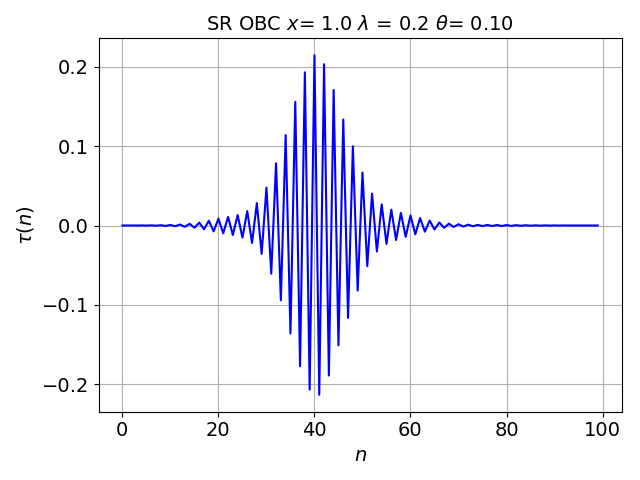}
	\caption{Same as Fig.\ref{fig: fig5} but with the symmetric regularization \label{fig: fig6}}
\end{figure*}

 The results obtained are plotted in Fig.\ref{fig: fig5} for the conventional regularization, and in Fig.\ref{fig: fig6} for the symmetric regularization. In both cases, we clearly observe that the excited state contains a {\it localized} excitation, the localization length of which grows smaller as the disorder strength is increased. Furthermore, for all parameter values that were considered, the localization length was longer than would be accessible using exact diagaonalization, necessitating DMRG. Finally, it is apparent from the figures that the elementary excitation is charge neutral, consistent with general expectations for the Schwinger model that excitations should be dipoles, which in one dimension lack long range interactions. We remained throughout in the regime where kinetic energy was the largest scale in the problem, and disorder the smallest, which is the regime where the analysis of Ref. \onlinecite{NandkishoreSondhi} is expected to apply. 

\par
\begin{figure}
	\centering
	(a)\includegraphics[width = 6cm]{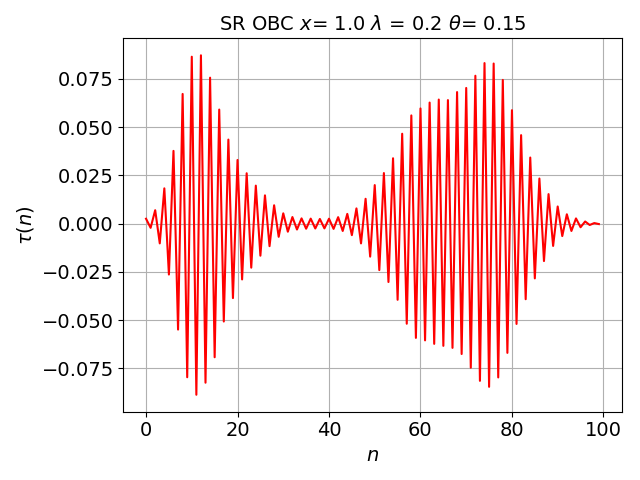}\\
	(b)\includegraphics[width = 6cm]{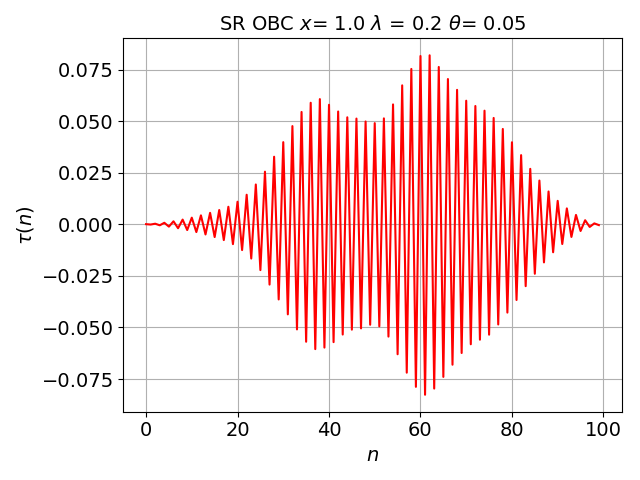}
	\caption{For certain disorder realizations, the density imbalance in the first excited state is bilocalized (a). In this case we take the localization length to be the mean of the decay lengths for the two peaks. In other cases, there are two overlapping peaks (b) which cannot be cleanly separated. We exclude such configurations from our estimate of localization length.}
	\label{fig: fig7}
\end{figure}

\begin{figure}
	\centering
	\includegraphics[width=6cm]{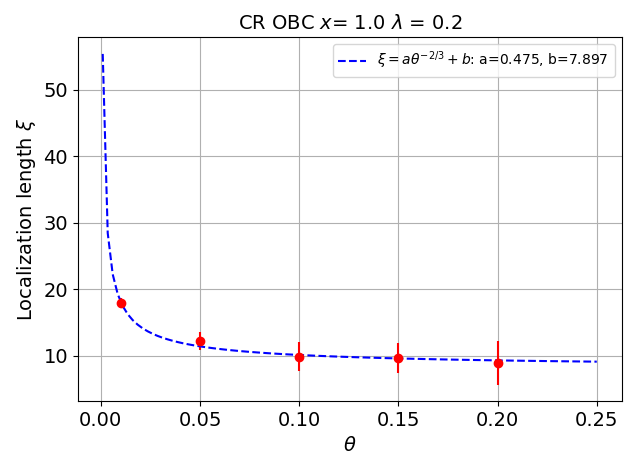}
	\includegraphics[width=6cm]{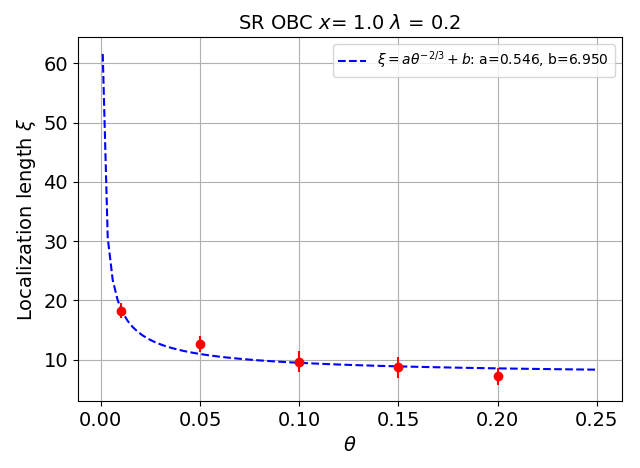}
	\caption{Localization length $\xi$ (extracted by eye from Fig.\ref{fig: fig5} and Fig.\ref{fig: fig6}), plotted against disorder strength. The dashed lines are a best fit to the scaling prediction $\xi \sim \theta^{-2/3}$. The error bars come from sample to sample fluctuations and are estimated by averaging over at least 10 disorder realizations.    \label{fig: fig8}}
\end{figure}

 Finally, it is interesting to explore how the localization length scales with disorder strength, and to compare to the analytical predictions in Ref. \onlinecite{NandkishoreSondhi}. The localization length can be read off from plots of the form Fig. \ref{fig: fig5}, \ref{fig: fig6}, and the error bars estimated by sampling multiple disorder realizations. However, there is a slight complication. For some disorder realizations, the `density imblance' turns out to be bimodal, corresponding to existence of a long range resonance in the system (see e.g. Fig.\ref{fig: fig7}a). In such cases, we take the localization length to be the mean of the decay lengths of the two peaks. In other cases, there are two peaks that overlap and cannot be cleanly separated (see e.g. Fig. \ref{fig: fig7}b). Such cases are excluded from our estimate of the localization length. The localization length obtained in this manner is plotted as a function of disorder strength in Fig. \ref{fig: fig8} for the two choices of regularization. Ref.\onlinecite{NandkishoreSondhi} predicted that the localization length $\xi$ should scale with disorder strength $\theta$ as $\xi \sim \theta^{-2/3}$. A fit to this prediction is in good agreement with the data.
 
\section{Discussion and conclusions}

\par We have explored the interplay of disorder, confinement and lattice commensuration in a lattice regularized version of the Schwinger model. We have pointed out that there are two distinct choices of lattice regularization which at first glance are very different. One is based on a picture of a two component plasma of electrons and positrons, and has a two site unit cell; while the other is based on a picture of electrons moving on a uniform positively charged background, and has a one site unit cell. Nonetheless, both regularizations have the same physics, because the `symmetric' regularization exhibits spontaneous symmetry breaking, such that the {\it eigenstates} of the Hamiltonian have a two site unit cell even though the Hamiltonian itself has only a one site unit cell. This spontaneous symmetry breaking survives the addition of disorder, with the Imry-Ma theorem being evaded due to the long range interaction. 

\par We have used exact diagonalization to examine the energy resolved level statistics ratio of the model, and have identified ergodicity breaking in the low energy sector of Hilbert space. However, in the system sizes to which exact diagaonlization is limited, we cannot determine whether this is due to localization or due to proximity to an integrable point. However, a DMRG investigation on systems with $N=100$ spins confirms that the first excited state contains a localized excitation, the localization length for which decreases with increasing disorder strength. This is in outline consistent with the picture of localization in the disordered Schwinger model advanced in Ref. \onlinecite{NandkishoreSondhi}. The analytical analysis also predicted a scaling relation for localization length $\xi$ as a function of disorder strength $\theta$ - as $\xi \sim \theta^{-2/3}$. Fig. \ref{fig: fig8} shows a fit of the data to this form and finds good agreement between the data and the prediction.

\par There remain a number of interesting open questions for future work. In this investigation, we restricted ourselves to a regime where kinetic energy is the largest energy scale in the problem, and disorder the weakest. This is the regime where the analytic treatment of Ref. \onlinecite{NandkishoreSondhi} is well controlled, and outside this regime there is no analytic understanding that we are aware of. However, dynamics reminiscent of MBL have also been numerically observed in a very different parameter regime \cite{Scardicchio} when the interaction strength is the largest energy scale in the problem. A systematic exploration of the Schwinger model in all its parameter regimes would be an interesting problem for future work, and may serve as a useful guide to the development of an analytical understanding. 

\par We note also that while we were able to confirm localization in the first excited state, in this model localization is expected to persist \cite{NandkishoreSondhi} even in states with a non-zero {\it density} of excitations. This expectation follows because (as is apparent from our numerics), the elementary excitations of the problem are charge neutral, and thus should lack long range interactions. Unfortunately the numerical techniques available to us were not able to tackle states with a finite density of excitations in system sizes large enough to see localization (while remaining in the parameter regime where an analytical understanding is available). Numerical advances will likely be necessary to be able to explore this highly excited regime, and this would also be a worthwhile problem for future work. 

\par Finally, while we have focused on the simplest model exhibiting confinement - the Schwinger model - confinement is a more general phenomenon. The interplay of disorder, confinement and lattice scale effects in more complicated (higher dimensional) settings would also be an interesting topic for future work. 

{\it Acknowledgements} We acknowledge useful conversations with Mari Carmen Banuls. A. A. Akhtar would especially like to thank her for an introduction to DMRG during a summer stay at MPQ. This material is based in part (RMN) upon work supported by the Air Force Office of Scientific Research under award number FA9550-17-1-0183. 

\bibliography{bib} 

\end{document}